\documentclass[12pt]{article}
\usepackage{epsfig}
\begin{document}
\begin{center}
{
{\Large {\bf The Simple Center Projection of $SU(2)$ Gauge Theory.} }

{
\vspace{1cm}
{B.L.G~Bakker$^a$, A.I.~Veselov$^b$, M.A.~Zubkov$^b$  }\\
\vspace{.5cm}
{ \it
$^a$ Department of Physics and Astronomy, Vrije Universiteit, Amsterdam,
The Netherlands \\
$^b$ ITEP, B.Cheremushkinskaya 25, Moscow, 117259, Russia 
}} }
\end{center}
\begin{abstract}
We consider the $SU(2)$  lattice gauge model. We propose a new gauge
invariant definition of center projection, which we call the Simple Center
Projection. We demonstrate  the center dominance, i.e., the coincidence
of the projected potential  with the full potential up to the mass
renormalization term at low energies.  We also consider the center
vortices and the center monopoles ({\em nexuses}). It turns out that
the behavior of such objects  qualitatively coincides with the behavior
of the vortices and monopoles in the Maximal Center gauge.  The
connection of the condensation of nexuses with the dual superconductor
theory is discussed.  Numerically the procedure of extracting the
center vortices proposed in this paper is much simpler than the usual
Maximal Center Projection.

\end{abstract}

\section{Introduction}

We may need to go a long way before we understand  how the confinement
mechanism works within nonabelian gauge theory. A lot of physicists
now consider the Abelian projection as the best way to understand the
confinement mechanism.

After Abelian projection one should consider the element of some 
Abelian subgroup of the gauge group instead of the full gauge group element. 
Thus, the  theory  becomes Abelian and one can look at the  picture of 
the confinement phenomenon in a simpler way.

Abelian projections differ from each other by the choice of the
subgroup of the gauge group and the projection method.  The closeness
of a given Abelian projection to the solution of the confinement
problem is measured as follows. Suppose that the link gauge group
elements $g_{\rm link}\in G$ are projected onto the elements of the
given Abelian subgroup $e_{\rm link}\in E \subset G$.  We consider
\begin{equation}
 Z_C = {\rm Tr} \prod_{\rm link\in C} e_{\rm link}
 \label{eq.010}
\end{equation}
instead of the Wilson loop and extract the potential from $Z_C$, (the
{\em projected} potential).  If that potential is close to the original
confining potential at sufficiently large  distances one can say that the
projection is suitable for the investigation of the confinement
mechanism.

Dealing with the $SU(2)$ gauge group, the Cartan subgroup $U(1)$ and the
center subgroup $Z_2$ have been considered. The most popular
projections are the Maximal Abelian and the Maximal Center
projections.  Those projections are achieved by minimization with
respect to the gauge transformations of the distance between the given
configuration of the link gauge field and the Cartan (center) subgroup
of $SU(2)$.  In both cases the potentials are very close to the $SU(2)$
confining potential but unfortunately do not exactly coincide with it.
Greensite et al. \cite{Gr2000,DFGGO1998,DFGO1998} have argued that only
central charges $q= \pm 1$ are confined in nonabelian gauge theories.

From the work of Bornyakov et al.  \cite{BKPV2000} it is known that gauge
fixing procedures suffer from a gauge copies problem. Several gauges
have been used so far in practical computations: the
direct and indirect \cite{DFGO1998} center gauges and
the Laplacian center gauge \cite{FE1999}. Only the first two suffers
from the occurrence of gauge copies, the last one is free from this
difficulty.  According to Refs.~\cite{Gr2000, FGO1999a} this problem disappears
for large lattices and when the number of Gribov copies considered is
increased.  Of course, this means that the computational effort must
also be increased.

In this work we propose a new center projection which is not connected
with partial gauge fixing. Thus, all the objects existing within the
projected theory have a gauge invariant nature. We call this procedure
the Simple Center Projection (SCP). Based on numerical simulations we
make the conjecture that the projected potential coincides with the
$SU(2)$ potential up to the mass renormalization term at sufficiently
large  distances.

Within the SCP  we can construct the center vortices and the center
monopoles (also known as {\em nexuses}, see Refs.~\cite{CPVZ1999} and
\cite{Vo1999}). The properties of the SCP monopoles found in our
investigations lead us to believe that those monopoles are the objects
which play the role of the Cooper pairs in the dual superconductor.

\section{Simple Center Projection}

We consider $SU(2)$ gluodynamics with  the Wilson action 
\begin{equation}
 S(U) = \beta
\sum_{\mathrm{plaq}} (1-1/2 \, \mathrm{Tr} U_{\mathrm{plaq}}).
 \label{eq.011}
\end{equation}
The sum runs over all the plaquettes of the lattice. The plaquette action 
$U_{\rm plaq}$ is defined in the standard way.

First we consider the plaquette variable
\begin{eqnarray}
 z_{\rm plaq} = 1 , & {\rm if} &  {\rm Tr} \, U_{\rm plaq} < 0, \nonumber\\
 z_{\rm plaq} = 0 , & {\rm if} &  {\rm Tr} \, U_{\rm plaq} > 0.
 \label{eq.020}
\end{eqnarray}
We can represent $z$ as the sum of a closed form $d N$ \footnote{We use
the formulation of differential forms on the lattice, as described for
instance in Ref.~\cite{PWZ1993}.} for $N\in \{0,1 \}$ and the form $2m +q$.
Here $N = N_{\rm link}$, $q\in \{0,1\}$, and $m\in {\sf Z} \!\! {\sf Z}$.

\begin{equation}
 z = d N + 2m +q .
 \label{eq.030}
\end{equation}
The physical variables depending  upon $z$ could be expressed through
\begin{equation}
 {\rm sign} \, {\rm Tr} \, U_{\rm plaq} = \cos (\pi (dN +q)) .
 \label{eq.040}
\end{equation}
We shall say that $N_{\rm link}$ is the center projected link
variable.  There are many different ways to make this projection. The
Maximal Center Projection (MCP) uses the gauge ambiguity to make all
link matrices $U$ as close as possible to $e^{i\pi N}$. Thus the 1-form
$N$ is fixed for every gauge configuration.

There exist several ways to achieve this \cite{DFGO1998}. One way, 
the {\em direct way}, is to minimize the quantity 
\begin{equation}
 R = \sum_x \sum_\mu {\rm Tr} [ U_\mu(x) ] {\rm Tr}  [U^\dag _\mu (x) ].
 \label{eq.041}
\end{equation}
In the {\em indirect way} one minimizes the quantity
\begin{equation}
 R' = 
 \sum_x \sum_\mu {\rm Tr} [U_\mu (x) \sigma_3 U^\dagger _\mu (x) \sigma_3 ],
 \label{eq.042}
\end{equation}
and extracts from $U_\mu(x)$ the diagonal part 
$A_\mu = \exp[i \theta_\mu(x) \sigma_3]$ thus fixing the gauge. Finally the
remnant Abelian symmetry is used to bring $A_\mu $ as close to an element of
$Z_2$ as possible by maximizing
\begin{equation}
 R'' = \sum_x \sum_\mu \cos^2 \theta_\mu(x).
 \label{eq.043}
\end{equation}
It is clear that both procedures are complicated and time consuming, as
the number of variables involved in the case of e.g. $SU(2)$ is three.
Whatever method is used, the 1-form $N$ is fixed for every gauge
configuration.

Now we choose a simpler and more natural procedure.  Imagine the
surface $\Sigma$ formed by the plaquettes dual to the "negative"
plaquettes (for which $z_{\rm plaq} = 1$). This surface has a boundary.
We enlarge the surface by adding a surface $\Sigma_{\rm add}$ so that:\\
1. the resulting surface $\Sigma^1=\Sigma+\Sigma_{\rm add}$ will be closed;\\
2. when we eliminate from the surface $\Sigma_{\rm add}$ the 
plaquettes carrying even numbers $z_{\rm plaq} = \dots, -4,\,-2,\,2,\,4, \dots$,
the area of the remnant surface will be minimal for the given boundary. 
 
So $\Sigma^1$ can be represented by the closed form $dN$ on the
original  lattice for the integer link variable $N \in \{0,1\}$. And $N$
is the required center projected link variable.

Numerically this procedure lookes as follows. For the given variable 
$z$ we should choose such a $Z_2$ variable $N$ that $[dN]\,{\rm mod}\, 2$ is 
as close as possible to $z$. It means that we minimize the functional

\begin{equation}
Q = \sum_{plaq} |(z - dN)\, {\rm mod}\, 2|
\end{equation}
with respect to $N$. 
  
The procedure works in the following way. We consider a given link $L$
and the sum over plaquettes

\begin{equation}
Q_{link} = \sum_{L \in {\rm plaq}} |(z-dN) \, {\rm mod} \,  2|
\end{equation} 

We minimize this sum with respect to the one link value $N$.  All links
are treated in this way and the procedure is iterated until a global
minimum is found.

We call this unique and simple procedure the Simple Center
Projection. Our projection method also finds local minima, Gribov copies,
but as the procedure is much simpler and faster that the methods use until
now, we can afford to repeat our calculations to include several Gribov
copies.

The physical meaning of the projected variables becomes clear after
considering the continuum limit. Naively, the considered surfaces
disappear as the field strength on them tends to infinity.
Nevertheless this fact should be investigated more carefully. In any
case for resonable sizes of lattices finite volume effects occur
and the scaling window ends at some value of $\beta$. Thus the
direct drop into the continuum limit is impossible and within the
scaling window our surfaces $\Sigma$ carry large but not infinite field
strength. The results of the next section give us the reason to believe
that these surfaces are those which play the crucial role in the
confinement mechanism.

Let us also mention that it is reasonable to consider the analogous
construction for which we change the plaquettes into loops of sizes
$2 \times 2$, $3 \times 3, ...$. These extended projections solve the
problem of the positive plaquette model, for which our "$1 \times 1$"
$\Sigma$ is absent. 

\section{Numerical results}
In the calculations we report on here, we used as our standard lattice
one with dimension $24^4$ and for investigations at finite temperature
one with size $24^3 \times 4$. For reasons of comparison we have
occasionally used smaller lattices, of sizes $12^3 \times 4$ and $16^3
\times 4$. Some results were checked on the larger lattice $32^3 \times
4$. We have mentioned the gauge copy problem. We checked that for
lattices with linear dimensions $L \leq 24$, which we have used, 15
copies are sufficient and so we used everywhere 15 Gribov copies.

\subsection{Center Dominance}
We consider the following definition of the projected Wilson loop
\begin{equation}
 W^{\rm SCP}_C = \frac{1}{8}\,  Z_C \, (3\pi/4)^{{\cal P}(C)/4},
 \label{eq.060}
\end{equation}
Where ${\cal P}(C)$ is the perimeter of the loop $C$.

It may be interesting for the reader that this expression is very
similar to the first term in the character expansion from
\cite{FGO1999a, Og1999}. One can get the above expression from that
formula substituting  the factor $1/8$ instead of $1/4$ and the power
${\cal P}(C)/4$ instead of  ${\cal P}$. However, as the authors of
Refs.~\cite{FGO1999a, Og1999} derive their expression for 
lattices without gauge fixing, they can derive their results with local
operators, using the characters of the gauge group to obtain dominance
of the fundamental representation. In our case, we perform gauge fixing, 
which results in a nonlocal operator. Consequently, we cannot use the same
derivation as Refs.~\cite{FGO1999a, Og1999}. As we have not been able to
find a rigorous derivation of Eq.~(\ref{eq.060}), one must consider our
results up till now as empirical ones.
It should be stressed that these results were checked for Wilson loops of
sizes up till $6 \times 6$. It might be useful to check whether 
Eq.~(\ref{eq.060}) continues to give good results if larger Wilson loops
are considered and the statistics is improved.

\begin{figure}[h]
\begin{center}
 \epsfig{figure=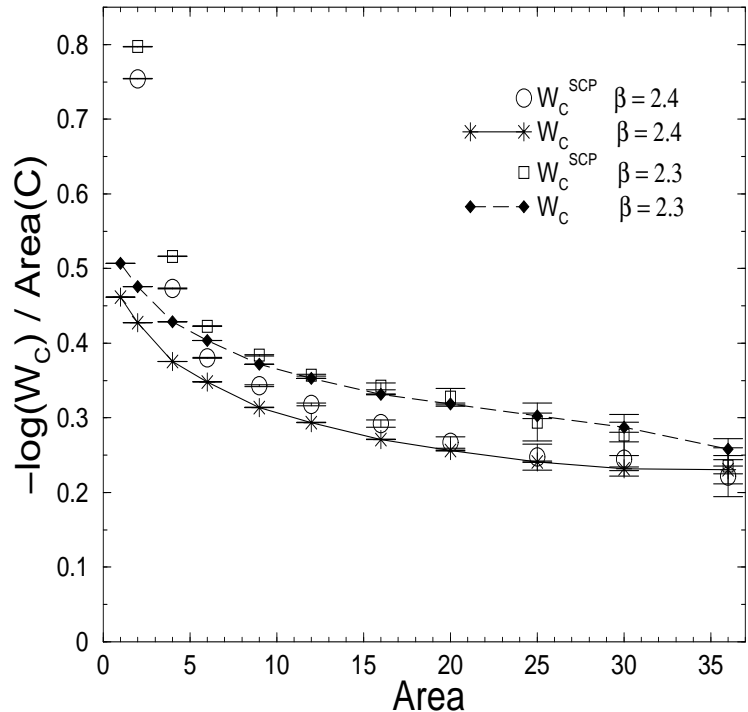,height=100mm,width=100mm}
 \caption{\label{fig.01} $-\log (W_C)/{\rm Area}(C)$ for all Wilson loops,
 $W_C$, and computed from projected ones, $W^{\rm SCP}_C$.
 $\beta$ = 2.3 and 2.4}
\end{center}
\end{figure}

The reader can recognize from Fig.~\ref{fig.01} that $ W_C $ and $
W^{\rm SCP}_C$ exactely coincide with each other for large enough sizes
of the loop (we represent here $-\log W_C /{{\rm Area}(C)}$ and $-\log
W^{\rm SCP}_C/{{\rm Area}(C)}$ as a function of the loop area for the
values $\beta$ = 2.3 and 2.4).
$W^{\rm SCP}_C$ differs from $Z_C$ by the perimeter factor. Thus we
understand that the projected potential (extracted from $Z_C$) differs
from the full potential (extracted from $W_C$) only by the  mass
renormalization term at low energies.  Other terms (including the
confining linear term) are the same.
It demonstrates the
Center Dominance.

\subsection{The center vortices}
We construct the closed two-dimensional center vortices as in \cite{CPVZ1999}
\begin{equation}
 \sigma = \, ^*d N .
\end{equation}
First, we can express $Z_C$ as 
\begin{equation}
Z_C = \exp(i\pi L\!\!L(\sigma,C))
\end{equation}
where $L\!\!L$ is the linking number \cite{CPVZ1999, CPV1997}.

Thus, according to the previous subsection the Aharonov - Bohm
interaction between the center vortices and the charged particle leads
to the confinement of the fundamental charge \cite{CPVZ1999, CPV1997}.

Also we investigate other properties of the center vortices for the finite 
temperature theory (the nonsymmetric lattice  $24^3 \times 4$). 

\begin{figure}[h]
\begin{center}
 \epsfig{figure=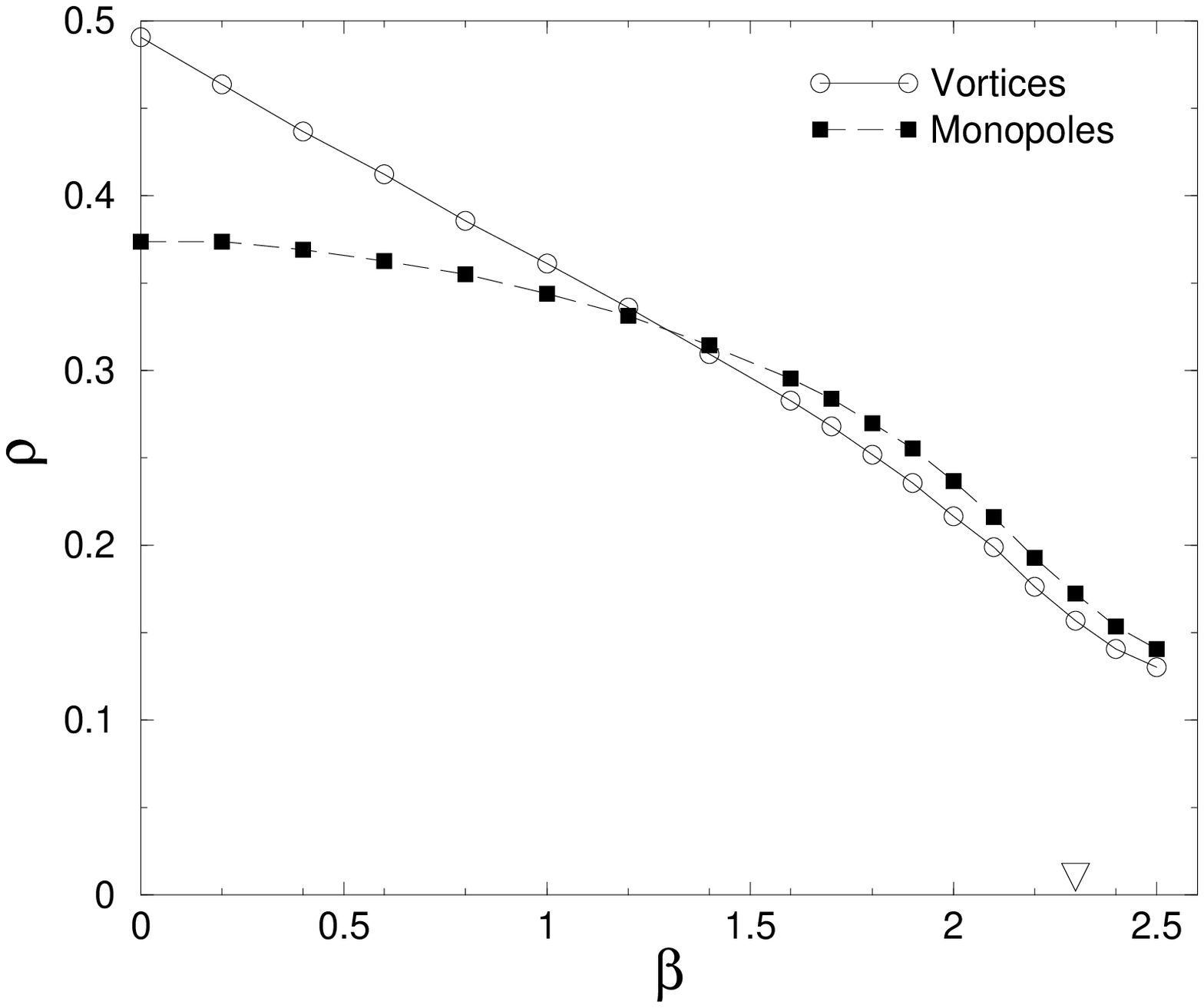,height=100mm,width=100mm}
 \caption{\label{fig.02} Density of center vortices and monopoles.
 The triangle points at the position of the phase transition.}
\end{center}
\end{figure}

\begin{figure}[h]
\begin{center}
 \epsfig{figure=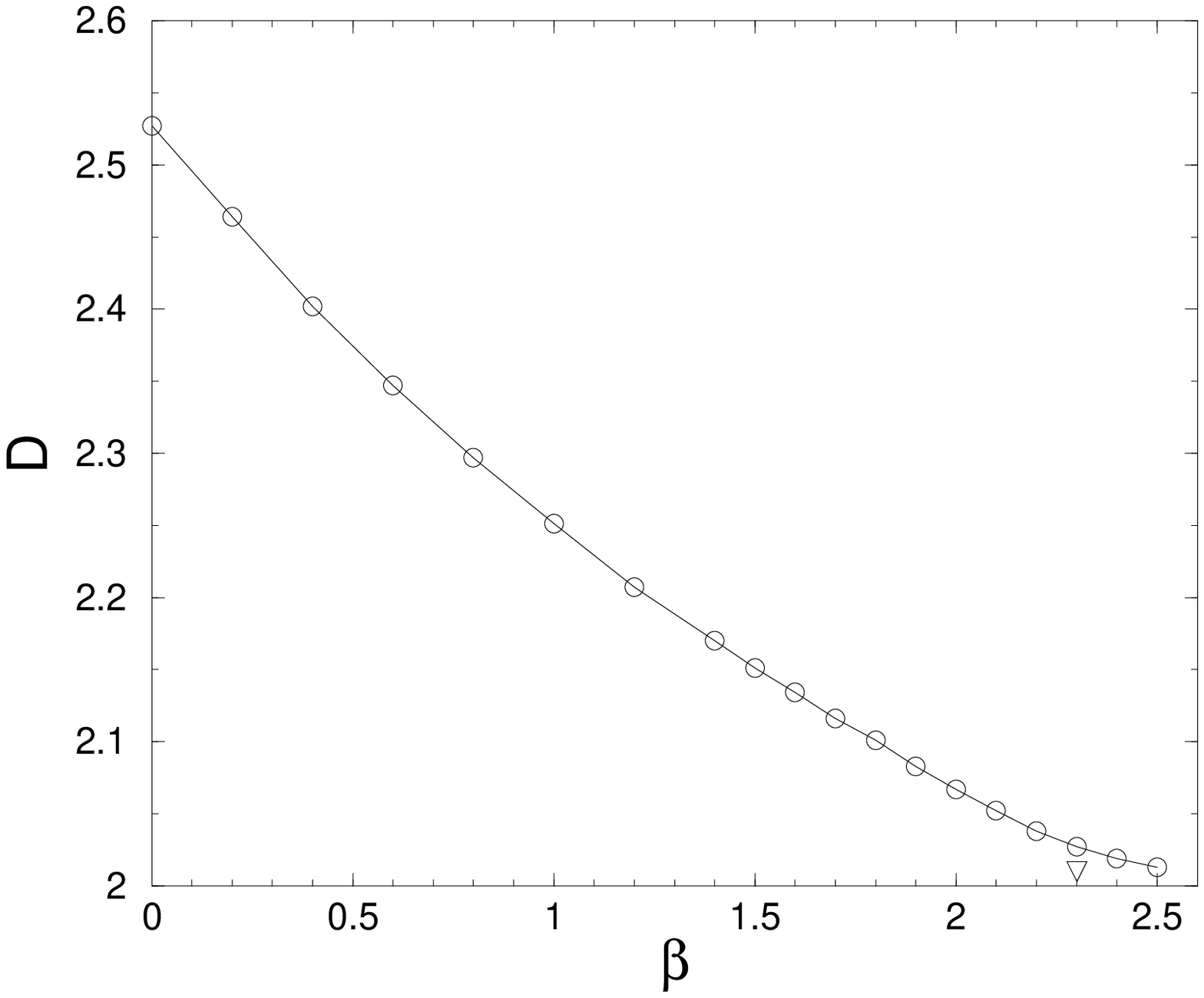,height=100mm,width=100mm}
 \caption{\label{fig.03} Fractal dimension $D$ of the center vortices.
 The triangle points at the position of the phase transition.}
\end{center}
\end{figure}

The density of the vortices $\rho$ is shown in Fig.~\ref{fig.02}.
The fractal dimension, defined as $D = 1 + 2 A/L$, where $A$ is the
number of plaquettes and $L$ is the number of links \cite{BVZ1999} of the
vortices, is shown in Fig.~\ref{fig.03}. A line $L$ is counted as
belonging to the vortex if at least one of the faces of a cube dual to
that link contains a plaquette with charge 1. 

\subsection{The center monopoles (nexuses)} 

Following \cite{CPVZ1999} we construct the center monopoles (nexuses)
\begin{equation}
 j = \frac{1}{2} \; ^*d [d N]\; {\rm mod}\, 2 .
 \label{eq.070}
\end{equation}
We show the density of the nexuses as a function of $\beta$ in 
Fig.~\ref{fig.02}.
It is important to make sure that the monopole lines are closed. It
follows from the equation $\delta j =\frac{1}{2} {}^*d{}^*{}^*d [dN]\; {\rm mod}\; 2 =0 $.

\begin{figure}[h] 
\begin{center}
 \epsfig{figure=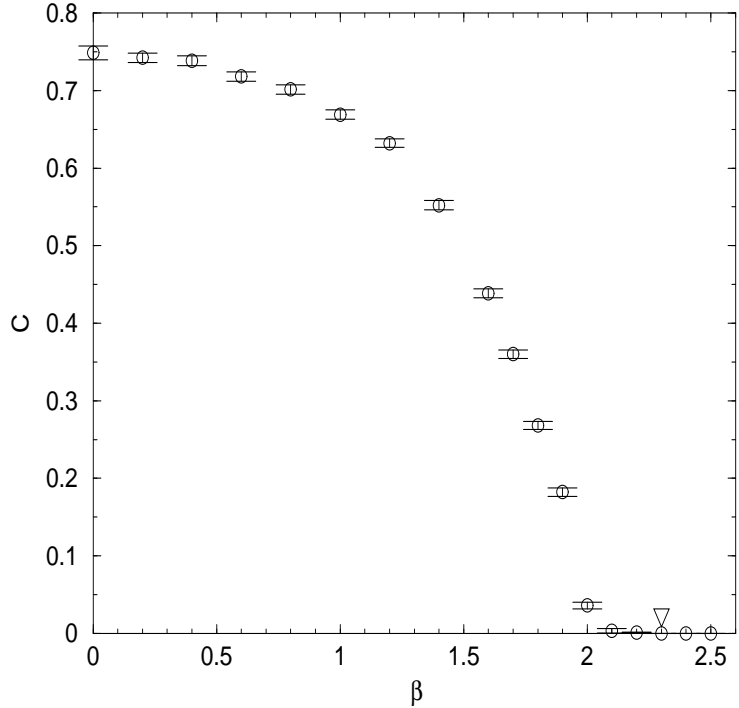,height=100mm,width=100mm}
 \caption{\label{fig.04} Probability of connecting two points by a
 monopole worldline for lattices $24^3 \times 4$.
 The triangle points at the position of the phase transition.} 
\end{center} 
\end{figure}

We investigated the percolation properties of $j$ and considered the
probability of two points to be connected by a monopole worldline on a
constant-time hypersurface. The dependence of that probability upon $\beta$ for
a nonsymmetric lattice is shown in Fig.~\ref{fig.04}.  The percolation
probability is dependent on the lattice size. This is obvious
for very small lattices, but it is seen to persists at larger sizes.

Here we would like to remark on the role the monopole condensate plays as an
order parameter. Ivanenko et al. \cite{IPP1993} have shown for abelian 
monopoles, obtained after Maximal Abelian projection, that the monopole
condensate can be used as an order parameter: it vanishes in the deconfined
phase, while it takes a finite value in the confining phase. The authors
of Ref.~\cite{IPP1993} made the interesting observation that the behavior of
the condensate near the point of the phase transition depends on the
``thickness'' of the monopole lines. 
It was demonstrated in Ref.~\cite{AGG1999} that Abelian monopoles are
strongly connected with vortices. For vortex lines with a thickness of
one lattice spacing, the condensate vanishes smoothly near the critical
point, whereas for lines with a thickness of two lattice units, the variation
of the condensate near the critical point is very steep.

We see that the monopoles are condensed in the confinement phase
and not condensed in the deconfinement phase, but the phase transition is
rather smooth as is the case for thin Abelian monopoles. 
If we would investigate, along the lines of Ref.~\cite{IPP1993} nexuses
of larger sizes like $2^3$, $3^3$ and so forth, we think that we shall
obtain a better sensitivity of the condensates $C^{(2)}$, $C^{(3)}$ etc.
to the phase transition. (Here we use the obvious notation $C^{(n)}$ for the
condensate of vortices of size $n^3$.

We expect the center monopoles to be
the monopoles that are present in the dual superconductor picture of
confinement. As the phase transition is not very pronounced for the
thin vortices we considered here, our results may be taken as a ``proof of
principle''. They must be substantiated by considering thick vortices which
are supposed to be more strongly connected to confinement \cite{FGO2000}.

The analytical connection of the monopole condensation and the
formation of the dual superconductor was considered in \cite{Zub} for
the case of $SU(3)$ symmetry. Of course, the results of that paper obtain
also for the $SU(2)$ theory.  It follows from \cite{Zub} that in
the case both condensation of nexuses and center
dominance occur, the picture of the dual superconductor in
which the nexuses play the role of Cooper pairs and the quarks play
the role of the monopoles becomes clear.

In particular, one can rewrite the fundamental Wilson loop as follows
\begin{eqnarray}
< W(C) > =\int_{-\pi}^{\pi} D H\int D_{\Phi\in C}\Phi
 \exp( -Q(dH+\pi^*A[C])\nonumber\\
 -\sum_{xy} \Phi_{x} e^{2iH_{xy}} \Phi_y^+ - V(|\Phi|))
\end{eqnarray}
Here $H$ is the electromagnetic field, $A[C]$ is the area of the
surface spanned on the quark loop, $\Phi$ is the nexus field, $Q$ is
nonlocal effective action, $V$ is an infinitely deep potential, supporting
the infinite value of the nexus condensate.

Thus we have indeed the nonlocal relativistic superconductor theory, in
which the nexuses are the Cooper pairs and the quarks are the monopoles.
The condensation of nexuses gives rise to the formation of the quark -
antiquark string appearing as the Abrikosov vortex.

\section{Conclusions}

In this work we construct the gauge invariant Center projection and
show that center dominance takes place. We investigate the properties
of the topological
defects in the center projected theory, which are shown to be closely connected
to the confinement picture. Particulary  it occurs that
the center monopoles from this projection are good candidates for
the Cooper pairs in the dual superconductor. The simple numerical nature
of the Simple Center Projection, the exact Center Dominance and the
properties of the topological defects existing in the center
projected theory give us the reason to propose SCP as the basic
abelian projection for the considering of the confinement picture.

\section*{Acknowledgments}

We are grateful to V.G. Bornyakov, M.N. Chernodub, and M.I. Polikarpov
for useful discussions.  A.I.V. and M.A.Z. kindly acknowledge the
hospitality of the Department of Physics and Astronomy of the Vrije
Universiteit, where part of this work was done.  This work was partly
supported by the grants RFBR 99-01-01230, INTAS 96-370.

\end{document}